\documentclass[aps,prx, twocolumn,amsmath, longbibliography, amssymb,superscriptaddress]{revtex4-2}

\usepackage{physics}
\usepackage{graphicx}% Include figure files
\usepackage{dcolumn}% Align table columns on decimal point
\usepackage{bm}% bold math
\usepackage{ulem}
\usepackage[dvipsnames]{xcolor}
\usepackage{amsmath}
\usepackage[colorlinks,allcolors=blue]{hyperref}

\newcommand{\one}{\mbox{$1 \hspace{-1.0mm}  {\bf l}$}}

\begin{document}

\preprint{APS/123-QED}

\title{Strategies for entanglement distribution in optical fiber networks}% Force line breaks with \\

\author{Hannah McAleese}
\affiliation{Centre for Quantum Materials and Technologies, School of Mathematics and Physics, Queen’s University Belfast, BT7 1NN Belfast, UK}

\author{Anuj Agrawal}
\affiliation{CONNECT Centre, Trinity College Dublin, Dublin, Ireland}

\author{Vivek Vasan}
\affiliation{CONNECT Centre, Trinity College Dublin, Dublin, Ireland}

\author{Conall J. Campbell}
\affiliation{Centre for Quantum Materials and Technologies, School of Mathematics and Physics, Queen’s University Belfast, BT7 1NN Belfast, UK}

\author{Adam G. Hawkins}
\affiliation{Centre for Quantum Materials and Technologies, School of Mathematics and Physics, Queen’s University Belfast, BT7 1NN Belfast, UK}

\author{Daniel C. Kilper}
\affiliation{CONNECT Centre, Trinity College Dublin, Dublin, Ireland}

\author{Mauro Paternostro}
\affiliation{Universit\`a degli Studi di Palermo, Dipartimento di Fisica e Chimica - Emilio Segr\`e, via Archirafi 36, I-90123 Palermo, Italy}
\affiliation{Centre for Quantum Materials and Technologies, School of Mathematics and Physics, Queen’s University Belfast, BT7 1NN Belfast, UK}

\author{Marco Ruffini}
\affiliation{CONNECT Centre, Trinity College Dublin, Dublin, Ireland}

\date{\today}% It is always \today, today,

\begin{abstract}
Distributing entanglement over long distances remains a challenge due to its fragility when exposed to environmental effects. In this work, we compare various entanglement distribution protocols in a realistic noisy fiber network. We focus specifically on two schemes that only require the sending of a non-entangled carrier photon to remote nodes of the network. These protocols rely on optical \textsc{cnot}/\textsc{cphase} gates and we vary the probability with which they can be successfully performed. Encoding our entangled states in photon polarization, we analyse the effect of depolarizing noise on the photonic states as the carrier passes through the fibers. Building a robust model of photon loss and calculating the distillable entanglement of the noisy states, we find the entanglement distribution rate. We discover that methods involving a separable carrier can reach a higher rate than the standard entanglement distribution protocol, provided that the success probability of the optical \textsc{cnot}/\textsc{cphase} gates is sufficiently high.
\end{abstract}

\maketitle

%\tableofcontents

\section{Introduction}

The distribution of entanglement across quantum communication channels opens the door to applications such as superdense coding~\cite{Bennett1992}, quantum teleportation~\cite{Bennett1993} and distributed quantum computing~\cite{distributedquantum}. This has led to the endeavor to create a {\it quantum internet}~\cite{Kimble2008,Wehner2018} allowing this advanced communication technology to connect remote stations across the globe. Part of the current efforts towards the achievement of this overarching goal is the identification of the best suited infrastructure to underpin a large-scale quantum communication network. 

As optical fiber networks are already widespread and well-established for classical communication, it is natural to ask how they can be adapted for quantum communication tasks addressing the short-to-medium haul scale (as genuinely global quantum communication might benefit from satellite-based links). The successful distribution of polarization-encoded entanglement in such networks has been demonstrated, which is a key milestone towards further developments~\cite{Wengerowsky2019,Shen2022,Neumann2022,Bersin2024,Craddock2024}. However, the only approach used so far has been direct entanglement distribution (DED), where an entangled pair of photons is produced and distributed to the desired nodes. However, photons will inevitably be exposed to noise, which would result in the eventual reduction or loss of any quantum correlations, thereby decreasing their usefulness for quantum applications. 

Alternative entanglement distribution protocols have thus been proposed, some of which do not require the sending or consumption of any entanglement at all~\cite{Cubitt2003,Mista2008,Mista2009,Kay2012,Mista2013,Fedrizzi2013,Vollmer2013,McAleese2021,campbell2024,Lavene2022,Chuan2012}. In this paper, we leverage such proposals and critically investigate their implementation in an optical fiber network, considering a realistic noisy environment, and using a standard DED method as a performance benchmark.   

To this end, we address quantitatively the DED scheme in the zero-added-loss multiplexing (ZALM) form~\cite{Chen2023}. We then address the paradigm of entanglement distribution via separable states (EDSS) through two protocols, both designed to bypass the use of any entangled resource in favor of weaker forms of quantum correlations (of the quantum discord form)~\cite{Ollivier2001,Henderson2001}, yet able to establish an entangled channel between two nodes of a network. 

One of the core achievements of our work is the identification of the required components and design of the node-architecture for the implementation of an EDSS scheme in an optical fiber network. The pursuit of such a goal allowed us to realize that the node architecture for EDSS is similar to the ZALM-based implementation of DED. A posteriori, this provides a justification for the choice of such an architecture as benchmark for DED in this work. Using the polarization degree of freedom to encode information, our quantitative assessment of performance includes the study of the effects of photon losses and fiber-induced depolarizing mechanisms on the polarization of the signals propagating across a network. 

A key part of our investigation is the comparative study of the protocols being addressed. If adopting the rate of entangled photons distributed by any given protocol as the quantitative figure of merit for efficiency, we find a stark dependence on the operating conditions of the network. In particular, the quality of the necessary logical gates in the EDSS schemes is pivotal: in general, EDSS can outperform DED as long as sufficiently high-quality gates are deployed, albeit at the cost of implementing suitable entanglement distillation procedures. Therefore, in quantum networks built with optical fibers, EDSS protocols could provide a route to achieving high rates of entanglement generation.

The remainder of this paper is structured as follows. In Sec.~\ref{sec:protocols}, we introduce the steps of each entanglement distribution protocol studied. In Sec.~\ref{sec:nodeArchitectures} we describe the different node architectures for DED and EDSS protocols. In Sec.~\ref{sec:loss} we study the probability of losing photons at different stages of each protocol. We take into account depolarizing noise in the optical fibers in Sec.~\ref{sec:noise}, analysing its impact on the entanglement of the photon pairs produced and their fidelity with a Bell state. In Sec.~\ref{sec:entGenerationRate} we distill the entanglement in the noisy photon pairs to calculate the entanglement distribution rate. Finally in Sec.~\ref{sec:conclusions}, we present our conclusions.

\section{Entanglement distribution protocols} \label{sec:protocols}

\begin{figure}[t!]
	\centering
	\includegraphics[width=\linewidth]{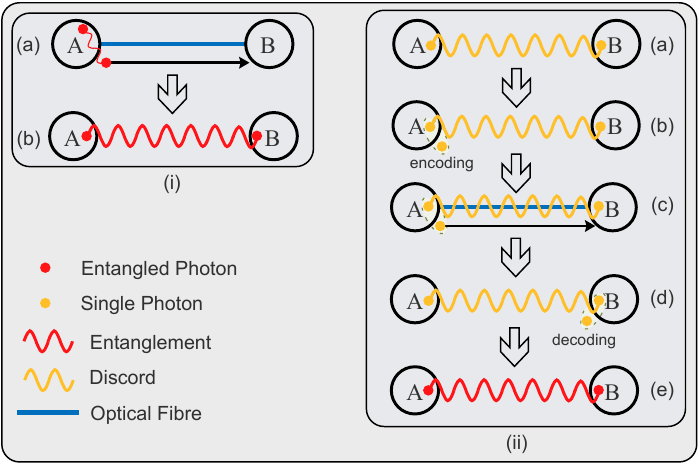}
	\caption{(i) Direct entanglement distribution (DED) versus (ii) entanglement distribution via separable states (EDSS).}
	\label{fig:dedvsdmed}
	\vspace{-0.5em}
\end{figure}

In this Section, we describe the two chosen EDSS protocols, which we label as EDSS1 and EDSS2. We focus solely on discrete-variable protocols, encoding the states in photon polarization. Fig.~\ref{fig:dedvsdmed} emphasizes the differences between DED and EDSS. For DED in Fig.~\ref{fig:dedvsdmed} (i), in order to establish entanglement between two remote nodes -- say nodes A and B -- an entangled photon pair is generated at node A and one of the entangled photons is sent to B, for instance via an optical fiber, while storing the other photon locally at node A. In the EDSS approach in Fig.~\ref{fig:dedvsdmed} (ii), discord is first established between nodes A and B by preparing single-photons at A and B in discordant states without the sending of any photons between the two nodes. Then a carrier photon is sent (after some encoding operation, detailed in Secs.~\ref{sec:EDSS1} and \ref{sec:EDSS2}) from one node to the other, followed by a decoding operation, to establish entanglement between the two nodes. In EDSS, the carrier photon remains separable from the two entangling systems but does still share quantum correlations, as explained later on. By avoiding the need to send entanglement across noisy channels, we hypothesize that EDSS could distribute Bell pairs to nodes A and B which are more robust to the effects of this noise than those produced using DED.  

Protocol EDSS1 is similar to the original EDSS protocol proposed by Cubitt et al.~\cite{Cubitt2003} where the carrier initially shares classical correlations with the two systems to be entangled. Protocol EDSS2 is one of a class of protocols proposed by Kay~\cite{Kay2012} which has been experimentally realised~\cite{Fedrizzi2013}. In the latter, the carrier initially shares no correlations with the other systems.

\subsection{EDSS1 Protocol} \label{sec:EDSS1}

\subsubsection{State preparation} \label{sec:initialStateEDSS1}

As illustrated in Fig.~\ref{fig:dedvsdmed} (ii), our first step (a) is the preparation of a discordant state of photons shared between nodes A and B. Calling $a$ and $c$ the photons at node A and $b$ the photon at node B, the initial state needed for EDSS1 is 
\begin{equation} \label{eq:discordState}
\begin{split}
    \rho &=  \frac{1}{6} \Big(\ketbra{HH}{HH}_{ab} + \ketbra{VV}{VV}_{ab} \Big) \otimes \ketbra{V}{V}_c \\ 
    & + \frac{1}{6} \left(\sum_{k=0}^3 \ketbra{\Psi_k}{\Psi_k}_a \otimes \ketbra{\Psi_{-k}}{\Psi_{-k}}_b\right) \otimes \ketbra{H}{H}_c ,
\end{split}
\end{equation}
where $\ket{\Psi_k} = \frac{1}{\sqrt{2}} (\ket{H} + e^{i k \pi/2} \ket{V})$. We have introduced the (anti-)diagonally polarized states of light  $\ket{\Psi_k}$ corresponding to $k=0$ ($k=2$) and the right (left) circular polarized ones associated with  $k=1$ ($k=3$). Therefore, in what follows we label $\ket{D} = \ket{\Psi_0}$, $\ket{R} = \ket{\Psi_1}$, $\ket{A} = \ket{\Psi_2}$ and $\ket{L} = \ket{\Psi_3}$. Details on how to prepare the state in Eq.~\eqref{eq:discordState} can be found in Appendix~\ref{appA}. Importantly, as no entanglement is present in the initial state, it can be produced using local operations and classical communication. No photons need to be sent across the network at this stage.

\subsubsection{Steps of the protocol} \label{sec:stepsEDSS1}

We call the second step of an EDSS protocol, shown in Fig.~\ref{fig:dedvsdmed} (ii) (b), the {\it encoding operation}. In this case, it takes the form of a \textsc{cnot} gate on photons $a$ and $c$ 
\begin{eqnarray}
    \textsc{cnot}_{ac} = \ketbra{H}{H}_a \otimes \one_c + \ketbra{V}{V}_a \otimes X_c
\end{eqnarray}
where $X = \left( \begin{array}{cc}
    0 & 1 \\
    1 & 0
\end{array} \right)$ is one of the Pauli matrices. After encoding, carrier $c$ shares quantum discord with photons $ab$. In step (c), the carrier photon $c$ is sent along the fiber to node B. The fourth step (d), or {\it decoding operation}, is another \textsc{cnot} gate on photons $b$ and $c$ where, as in the encoding operation, $c$ is the target. At this point, A and B share entangled photons, but this entanglement is not maximal. The density matrix of $abc$ at this point is
\begin{equation} \label{eq:edss1rhoprime}
    \rho' = \frac{1}{3} \ket{\phi^+}\bra{\phi^+}_{ab} \otimes \ket{H}\bra{H}_c + \frac{2}{3} \one_{ab} \otimes \ket{V}\bra{V}_c,
\end{equation}
where $\ket{\phi^+} = \frac{1}{\sqrt{2}} (\ket{HH}+\ket{VV})$ is a Bell state. As we aim to create such maximally entangled states, we add an additional step and  measure the polarization of the carrier in the horizontal/vertical basis to increase the entanglement between A and B. As can be seen in Eq.~\eqref{eq:edss1rhoprime}, we get the outcome corresponding to horizontal polarization with probability 1/3, and in this case A and B share a Bell pair. If it is vertically polarized, however, we obtain a maximally mixed state and the correlations between A and B are destroyed. 

\subsection{EDSS2 Protocol} \label{sec:EDSS2}

\subsubsection{State preparation}
\label{sec:initialStateEDSS2}

In EDSS2, the carrier does not share any correlations, classical or quantum, with the photons we wish to entangle at the beginning of the protocol. We require photons $a$ and $b$ to be in state 
\begin{equation} \label{eq:initialEDSS1}
\begin{split}
    \rho_{ab} = & \frac{1}{4} (\ketbra{HH}{HH} + \ketbra{VV}{VV} )_{ab} +\frac{1}{8} (\ketbra{DD}{DD} \\ & + \ketbra{AA}{AA} + \ketbra{RL}{RL} + \ketbra{LR}{LR})_{ab}
\end{split}
\end{equation}
and carrier $c$ to be in state
\begin{equation}
    \rho_c = \frac{1}{4} \left(\ketbra{D}{D} + {3} \ketbra{A}{A}\right)_c. 
\end{equation}

\subsubsection{Steps of the protocol} \label{sec:stepsEDSS2}

EDSS2 follows the same steps as EDSS1, albeit some differences exist between the two schemes. Firstly, the encoding and decoding operations are implemented by the \textsc{cphase} gates
\begin{equation}
    \textsc{cphase}_{xc} = \ketbra{H}{H}_x \otimes \one_c + \ketbra{V}{V}_x \otimes Z_c,
\end{equation}
where $Z = \left( \begin{array}{cc}
    1 & 0 \\
    0 & -1
\end{array} \right)$ is another Pauli matrix and $x=a$ ($x=b$) for the encoding (decoding) step. After decoding, we can write the density matrix of $abc$ as
\begin{equation} \label{eq:edss2rhoprime}
    \rho'=\frac{5}{8} \sigma^1_{ab}\otimes \ket{A}\bra{A}_c + \frac{3}{8} \sigma^2_{ab} \otimes \ket{D}\bra{D}_c,
\end{equation}
where $\sigma^{1,2}$ are two Bell-diagonal states taking the form
\begin{equation}
\begin{split}
    & \sigma = p_1 \ket{\phi^+}\bra{\phi^+} + p_2 \ket{\phi^-}\bra{\phi^-} \\ & + p_3 \ket{\psi^+}\bra{\psi^+} + p_4 \ket{\psi^-}\bra{\psi^-},
\end{split}
\end{equation}
with $\ket{\phi^-} = \frac{1}{\sqrt{2}} (\ket{HH}-\ket{VV})$ and $\ket{\psi^\pm} = \frac{1}{\sqrt{2}} (\ket{HV}\pm\ket{VH})$ the remaining Bell states and where $\{p_i\}_{i=1}^4$ are non-negative probabilities satisfying $\sum_{i} p_i=1$. State $\sigma^1$ has coefficients $p_1 = 0.6$, $p_2 = 0.3$ and $p_3=p_4=0.05$ and is therefore entangled (since the maximum coefficient $p_1>0.5$). In contrast, state $\sigma_2$ is separable as its coefficients ($p_1=1/3$, $p_2=1/6$ and $p_3=p_4=0.25$) are all less than 0.5. Secondly, in the final step we need to use the $\{\ket{D},\ket{A}\}$ measurement basis. Outcome $\ket{A}$ is needed for entanglement to be successfully distributed and the probability of obtaining this is 5/8 as shown in Eq.~\eqref{eq:edss2rhoprime}. It is worth noting that the success probability of EDSS2 is higher than that of EDSS1. However, we cannot produce maximally entangled states using EDSS2. Throughout this paper, in order to quantify entanglement, we will use negativity~\cite{Vidal2002}
\begin{equation}
    N(\rho_{AB}) = \frac{\| \rho_{AB}^{T_A} \|_1 - 1}{2},
\end{equation}
where $\rho^{T_A}$ is the partial transposition of the density matrix $\rho$ with respect to subsystem $A$ and $\| \alpha \|_1 = \mathrm{Tr} \sqrt{\alpha^\dagger \alpha}$ is the trace norm of the arbitrary matrix $\alpha$. The largest negativity that can be achieved using the EDSS2 protocol is only 0.1 (compared to 0.5 for a Bell state). If we require Bell states, we would need to carry out entanglement distillation.

\section{Node architectures} \label{sec:nodeArchitectures}

\begin{figure}[t!]
    \centering
    \includegraphics[width=0.9\columnwidth]{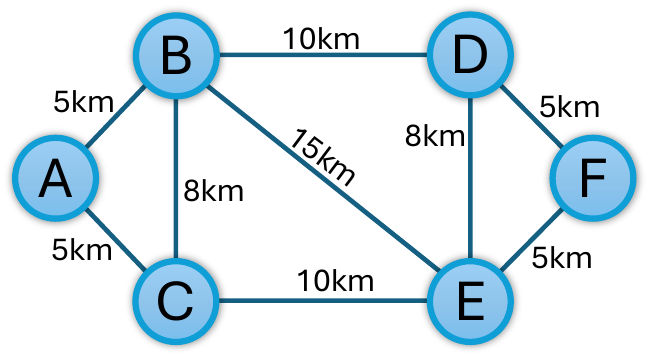}
    \caption{An example six-node mesh network used to analyze DED and the EDSS-based protocols. We show the fiber link distances within the considered network.}
    \label{fig:network}
\end{figure}

\begin{figure}[t!]
	\centering
	\includegraphics[width=\linewidth]{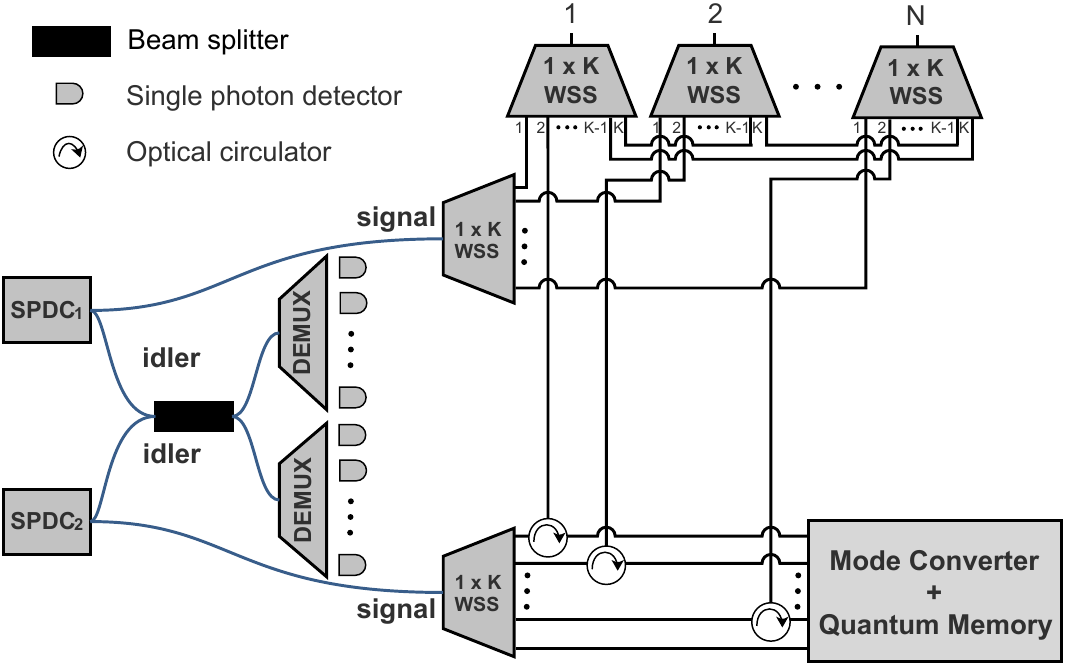}
	\caption{Node architecture for direct entanglement distribution (DED) in a network. Here, $\text{SPDC}_{1,2}$ stands for the {spontaneous parametric down-conversion} settings responsible for the generation of EPR pairs. WSS stands for wavelength-selective switch. The scheme uses a dual-SPDC-based heralded entanglement source.}
	\label{fig:dedsource}
\end{figure}

\begin{figure*}[t!]
	\centering
	\includegraphics[width=0.8\linewidth]{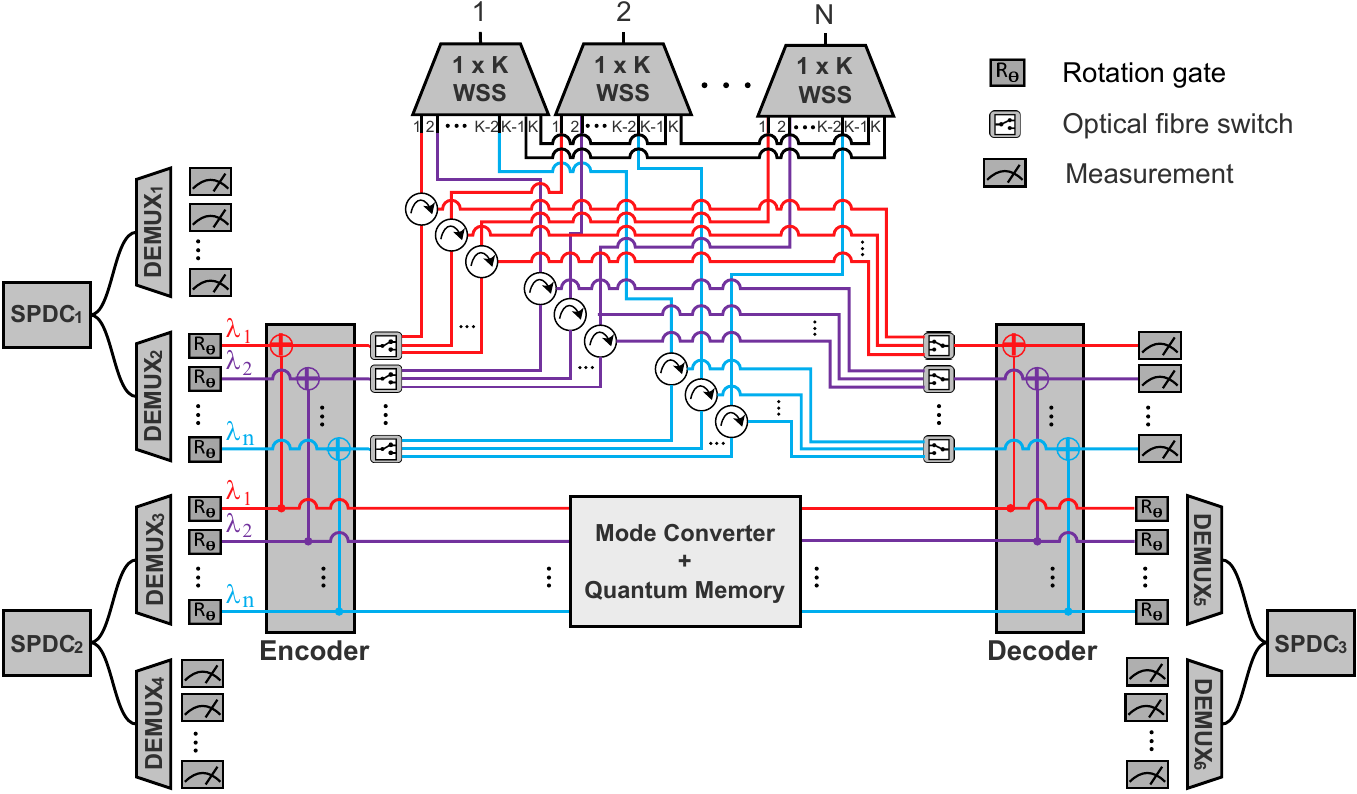}
	\caption{Node architecture for EDSS across a network. Here, $\text{DEMUX}_j$ is the $j^\text{th}$ demultiplexer. Other symbols and acronyms as in Fig.~\ref{fig:dedsource}. The \textsc{cnot} gates in the encoder and decoder should be replaced by \textsc{cphase} gates for EDSS2 protocol.}
	\label{fig:dmedsource}
\end{figure*}

We aim to use the three protocols in Sec.~\ref{sec:protocols} to distribute bipartite entanglement between each pair of nodes in the six-node network illustrated in Fig.~\ref{fig:network}. This mesh-connected network topology enables us to study entanglement distribution across a range of node pair connections; point-to-point and multihop (without and with intermediate node/s respectively between the two nodes we aim to entangle). The link distances are sufficiently short to avoid the need for quantum repeaters. While point-to-point entanglement distribution is straightforward, distributing entanglement between physically unconnected nodes in an optical network via intermediate nodes is not only complex but also requires additional switching components at nodes to add, drop, or bypass the photons. Thus, the add/drop/bypass functionality introduces additional loss in the system due to the switching components. We present node architectures for DED and EDSS schemes in Fig.~\ref{fig:dedsource} and Fig.~\ref{fig:dmedsource}, respectively. These architectures facilitate the following capabilities in the network: (i) any node in the network can be used as an entangled photon pair source node, photon receiving/quantum memory node, and/or intermediate node, (ii) wavelength division multiplexing (WDM)-based simultaneous entanglement distribution to multiple node pairs in the network using any (single) source node by routing different wavelengths to different node pairs (e.g., in Ref. \cite{Wengerowsky2019}) and (iii) simultaneous operation of more than one node in the network as source node (though this introduces congestion on fiber links). In this work, we consider entangling one node pair in the network at a time, where one of the entangling nodes operates as a source node and the other as the receiving node. We do not perform simultaneous entanglement distribution to multiple node pairs from single source or simultaneous operation of more than one node as source node as these operations introduce network-level issues that require efficient routing and spectrum allocation schemes. Moreover, the results in such modes of network operation will vary with network topology and will restrict our comparisons of the DED and EDSS protocols, which is the main focus of this work. Thus, we plan to perform an analysis on networking aspects utilizing the full features of these node architectures in our future work.

We follow the ZALM approach of entanglement distribution~\cite{Chen2023} in the DED case. In this approach, a dual-spontaneous parametric down-conversion (SPDC) setup is used to generate quasi-deterministic time-frequency heralded Bell pairs. Each SPDC source emits a signal photon and an idler photon, where the frequency of the signal is greater than or equal to that of the idler. Here, the idea is to perform frequency-demultiplexed entanglement swapping using the idlers of the two SPDCs, which results in entanglement between the two signals, as shown in Fig.~\ref{fig:dedsource}. Both the entangled signal outputs pass through a wavelength selective switch (WSS), which allows us to select a part of the spectrum (or subset of frequencies) on any of the WSS output ports. The array of WSSs on top in Fig.~\ref{fig:dedsource} is used to perform add/drop/bypass of photons. For instance, the signal from SPDC$_1$, after passing through the first WSS, can be routed out from the source node through any of the WSSs on top (depending on the location of the other entangling node in the network). When a node functions as an intermediate node, the incoming photons can be bypassed using the array of interconnected WSSs. For instance, a photon enters from the second WSS and exits from the first WSS of the array. When a node is the receiving node, a photon can enter from one of the top WSSs and is routed to the mode converter and quantum memory using the optical circulator. The other signal from SPDC$_2$ is stored in the quantum memory within the source node (notice that one of the output ports of the bottom WSS is directly connected to the mode converter and quantum memory). When the source node wants to send both signals out towards two other nodes in the network, the other ports of this WSS (except the bottom port) are used. However, as we entangle one node pair at a time in this work, one of the two signals is always stored locally within the node and the other is sent to the second node of the pair. The node architecture shown in Fig.~\ref{fig:dedsource} is adapted from Ref.~\cite{jsac2025} and is modified to (i) enable a node to either send or receive photons using the added optical circulators and (ii) represent a generalized node of degree N using an array of N interconnected WSSs.

We then create a new EDSS node architecture, shown in Fig.~\ref{fig:dmedsource}, adapting these protocols to optical fiber networks with wavelength division multiplexing for the first time. Here, three SPDCs are used: SPDC$_2$ and SPDC$_3$ generate the two photons to be entangled (i.e., photons $a$ and $b$ in Section \ref{sec:EDSS1}) and SPDC$_1$ generates the carrier photon $c$. SPDC$_1$ and SPDC$_2$ operate when a node acts as the source node and SPDC$_3$ operates when a node acts as a receiving node in the network. The idlers of SPDC$_1$ and SPDC$_2$ are used to herald the initial state of the signal photons. The encoding operation is performed between the same-wavelength signal photons from SPDC$_1$ and SPDC$_2$. After the encoding operation, the carrier photons (from SPDC$_1$) are sent out of the node towards the other entangling node and the photons from SPDC$_2$ are directed towards the mode converter and quantum memory present within the node. The function of the interconnected WSS-array is the same as that in the node architecture for DED, i.e., to add/drop/bypass the photons at a node. The optical fiber switches facilitate the routing of carrier photons to/from any of the WSSs in the WSS-array. When the node operates in the receiving mode, the incoming photons are routed through the optical circulators towards the decoder, where after decoding, the carrier photons are measured and the photons from SPDC$_3$ are directed towards the mode converter and quantum memory. If the measurement outcome for a carrier photon is as desired by the EDSS protocol, the corresponding photons with which the carrier photon underwent encoding and decoding become entangled. An important difference to notice between the DED and EDSS node architectures is that in EDSS, the photons to be entangled are never sent out of the node whereas in DED the entangled photons are sent out from the source node and traverse the network.

\section{Effect of photon losses on the performance of the protocols}
\label{sec:loss}

We now study the lossy components of the networks described in Sec.~\ref{sec:nodeArchitectures}. Due to the fiber loss, the probability to lose any photon is relatively high; if we consider a typical loss coefficient of 0.2 dB/km, half the photons are lost every 15 km. In addition, the switching nodes employ WSSs, which are also lossy (as explained below). In all three protocols, should we lose any one photon at any point, we must recommence the protocol from the beginning. This is due to the correlations required between the photons; tracing out one subsystem of $ab$ (for DED) or $abc$ (for EDSS) spoils the correlations between the remaining photons and we must therefore prepare a new initial state.

For DED, the photon loss is modeled as a Bernoulli process with mean number of successful events 
\begin{equation}
\label{eq1}
E[S]=np.
\end{equation}
We consider an event as successful when both the entangled photons of an Einstein–Podolsky–Rosen (EPR) pair reach their respective entangling nodes. The pair generation rate of SPDC is denoted $n$. The probability of success $p$ is given by
\begin{equation}
\label{eq7}
p=p_{\text{succ}}(S_1) p_{\text{succ}}(S_2) p_{\text{succ}}(\text{swap}), 
\end{equation}
where 
\begin{equation}
\label{eq9}
p_{\text{succ}}(\text{swap})=\frac{\eta_{\mathrm{coup}}^{2} 
  p_{\mathrm{bsm}}}{L_{\mathrm{demux}}^2}
\end{equation}
is the probability of successful entanglement swapping between the two EPR pairs generated by two SPDCs. Here, $\eta_{\mathrm{coup}}$ is the efficiency of coupling the photons from free-space to fiber, $p_{\mathrm{bsm}}$ is the probability of successful Bell state measurement, and $L_{\mathrm{demux}}$ is the demultiplexer (shown in Fig. \ref{fig:dedsource}) loss. The probability of success of a signal photon ($S_1$) reaching one of the two entangling nodes is obtained as 
\begin{equation}
\label{eq10}
p_{\text{succ}}(S_1)=\frac{\eta_{\mathrm{coup}}}{L_{\mathrm{wss}}^3 \alpha_{\mathrm{lin}}^{D_{P1}} L_{\mathrm{node}}^{N_{P1}}},
\end{equation}
where $L_{\mathrm{wss}}$ is the loss of WSS, $\alpha_{\mathrm{lin}}$ is the fiber attenuation coefficient (in linear scale), $D_{P1}$ is the length of fiber on path 1 (from entanglement source to one of the two entangling nodes) and $L_{\mathrm{node}}$ is the loss due to an intermediate node. The term $L_{\mathrm{wss}}^3$ is due to the two WSSs used when a photon is sent out from the source node, as shown in Fig.~\ref{fig:dedsource}, and one WSS used when an incoming photon at the receiving node is directed towards the mode converter and quantum memory. In both the DED and EDSS cases, an intermediate node as a bypass involves two WSSs so that $L_{\mathrm{node}}=L_{\mathrm{wss}}^2$. $N_{P1}$ denotes the number of intermediate nodes on path 1. The values we take for these loss parameters can be found in Table~\ref{tab:parameters}.

The probability of success of a signal photon ($S_2$) reaching the internal memory of the source node is given by
\begin{equation}
\label{eq11}
p_{\text{succ}}(S_2)=\frac{\eta_{\mathrm{coup}}}{L_{\mathrm{wss}}},
\end{equation}
as evident from Fig. \ref{fig:dedsource}. We assume that the loss of optical circulators and beam splitters is minimal and hence it is neglected in this work. We also assume that there is no time delay between the idlers interfering at the beam splitter and that the filtering using DEMUXes is perfect.

Though we take into account the photon loss of the individual components of the ZALM source, it is worth highlighting that we do not carry out an in-depth analysis of the source and so we are not including other losses which appear in the dual-SPDC heralded source and the Bell state measurement; we therefore present a somewhat idealised version of ZALM. The implementation and performance of a dual-SPDC heralded entanglement source depend on various factors, such as the phase matching conditions of SPDC, the optical interferometric setup used and the pumping of the crystal. One of the possible implementations of ZALM has been recently studied in detail in Ref.~\cite{Shapiro2024} using two Sagnac interferometer configured SPDCs with type-0 degenerate phase matching conditions, each arranged to emit Bell singlet states. This study suggests that the rates could in fact be considerably lower than those used in our analysis. Our results should therefore be taken as an upper limit of the pair rates that can be achieved.

\begin{table}[]
\centering
\begin{tabular}{c|c}
    Parameter & Value \\ \hline 
    $n$ & $2.87 \times 10^7$ pairs/s~\cite{Cheng2019} \\
    $\eta_{\mathrm{coup}}$ & $85\%$~\cite{Guerreiro2013} \\
    $p_{\mathrm{bsm}}$ & 0.5 \\
    $L_{\mathrm{demux}}$ & 3 dB \\
    $L_{\mathrm{wss}}$ & 3.5 dB~\cite{Fontaine2018} \\
    $\alpha$ & 0.17 dB/km \\
    $L_{\mathrm{sw}}$ & 1 dB \\ 
\end{tabular}
\caption{Values used for the parameters in Sec.~\ref{sec:loss}. Those given in units of decibels must be converted to linear scale before calculating rates and success probabilities in Eqs.~\eqref{eq1}-\eqref{eq13}.}
\label{tab:parameters}
\end{table}

Moving on to the EDSS case, the photon loss model does not vary with the protocol used. Moreover, unlike DED, an entangled photon pair is not generated at the source in this case, which simplifies loss modeling. Assuming the same pair generation rate $n$ of all the SPDCs used in DED (cf. Fig.~\ref{fig:dedsource}) and EDSS (cf. Fig.~\ref{fig:dmedsource}), let $n_{\lambda_{i}}$ be the photon pair generation rate of SPDC at wavelength $\lambda_{i}$. The photon output rate of the $j^{th}$ demultiplexing unit (DEMUX) at $\lambda_{i}$ is obtained as 
\begin{equation}
\label{eq12}
n_{\lambda_{i},d_j}=\frac{n_{\lambda_{i}}\eta_{\mathrm{coup}}}{L_{d_j}},
\end{equation}
where $L_{d_j}$ is the loss of the $j^{th}$ DEMUX. The EDSS pair distribution rate is calculated as
\begin{equation}
\label{eq13}
R_{\lambda_{i}}=\frac{\min_j(n_{\lambda_{i},d_j}) P^2 p_{\mathrm{meas}}}{L_{\mathrm{sw}}^2 L_{\mathrm{wss}}^2 L_{\mathrm{node}}^N \alpha_{\mathrm{lin}}^D},
\end{equation}
where $P$ is the success probability of the encoding and decoding operations, $p_{\mathrm{meas}}$ is the probability of obtaining the needed outcome when measuring the carrier photon at the end of the protocol, $L_{\mathrm{sw}}$ is the loss of fiber switch, $D$ is the length of fiber between the two nodes being entangled, and $N$ is the number of intermediate nodes between the two nodes being entangled. In our simulations, we assume $L_{d_j}=L_\mathrm{demux} ~\forall j$.

\begin{figure*}[t]
    \centering
    {\bf (a)}\hskip9cm{\bf (b)}\\
    \includegraphics[width=\columnwidth]{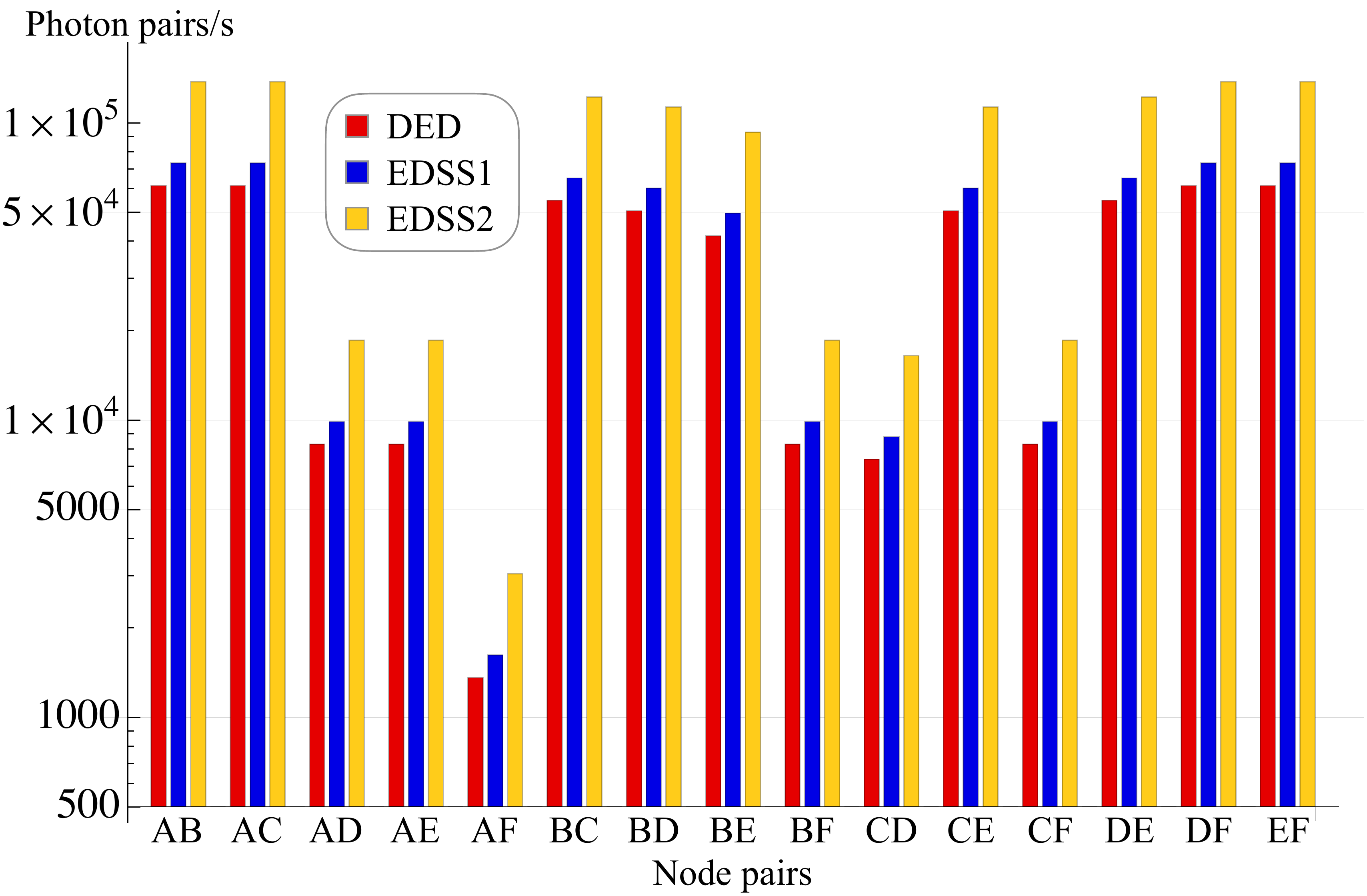}    \includegraphics[width=\columnwidth]{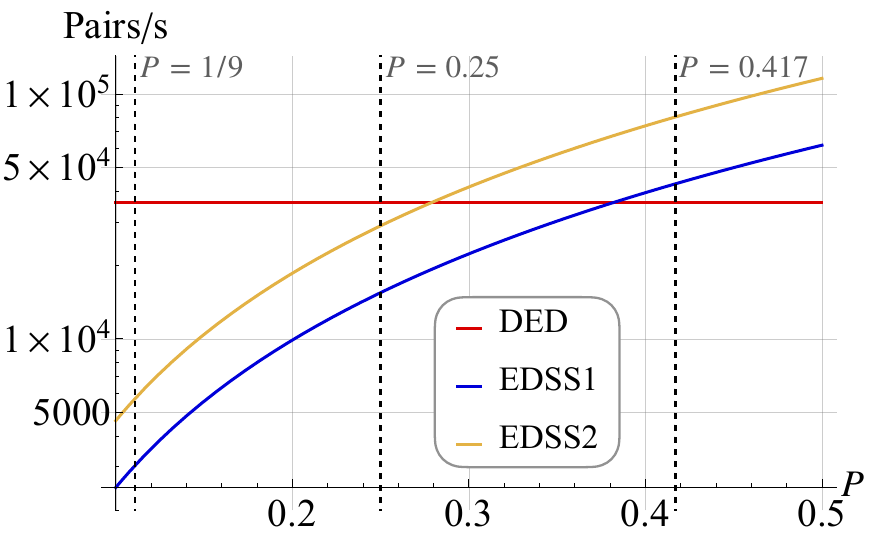}
    \caption{{\bf (a)} Rate of entangled photon pairs distributed  when considering losses in the nodes and transmission fibers. Note that for EDSS2, the state of the distributed photon pair is not maximally entangled. {\bf (b)} Average photon pair distribution rate over all the node pairs as the success probability $P$ of the encoding operation increases. The vertical black dashed lines highlight the typical experimental value for the success probability of an all-optical \textsc{cnot} gate~\cite{OBrien2003,Politi2008,Bao2007,Li2021,Pittman2003,Gasparoni2004,Stolz2022}.}
    \label{fig:pairDistributionRates}
\end{figure*}

So far, several experimental methods have been proposed for optical \textsc{cnot} or \textsc{cphase} gates. Most have a success probability of $\simeq11\%$ (cf. Refs.~\cite{OBrien2003,Politi2008}), which can grow to $12.5\%$~\cite{Bao2007,Li2021} and $25\%$~\cite{Pittman2003,Gasparoni2004}. A promising setup was recently put forward that achieved an average success probability of $41.7\%$~\cite{Stolz2022}. We will therefore vary this success probability $P$ of the encoding and decoding operations to analyse its impact on the final entanglement distribution rate. It is, however, important to highlight that we are taking a high-level view of \textsc{cnot} gate implementations and that many have practical limitations which could limit their use in our scheme. For instance, the method in Ref.~\cite{Stolz2022} requires a Rydberg atom stored at cryogenic temperatures and a narrow photon bandwidth due to atom-photon interaction, Ref.~\cite{Pittman2003} features destructive photon measurements, and in Ref.~\cite{Gasparoni2004}, the fidelity of the final state with its target is reduced due to birefringence in the polarizing beam splitters. We therefore take these success probability values as mere benchmarks. Moreover, as this is an active area of research, we expect new and improved methods to be developed in future.

The effect of loss on the number of entangled photons which can be distributed between each pair of nodes is presented in Fig.~\ref{fig:pairDistributionRates} {\bf (a)}. In this instance, we take the probability of successfully performing a \textsc{cnot} gate to be 0.417. It is useful to notice that many pairs have no intermediate nodes and could therefore be equivalently viewed as two-node networks; for example, node pair AB can be seen as a simple network of two nodes situated 5km apart. While EDSS2 consistently produces the highest number of entangled pairs, it is important to note that it does not produce maximally entangled photons in contrast to DED and EDSS1. Conversely, the pair distribution rate is lowest for DED. This may seem surprising, given the probabilistic nature of EDSS protocols which require specific measurement outcomes for success. However, DED features a Bell measurement on the idler photons of two photon pairs and the necessary outcome of this measurement is only obtained with a probability of 1/2. Moreover, the DED architecture requires two extra WSS components than EDSS, which can instead use optical fiber switches with less than a third of the loss. It must nevertheless be stated at this point that the results for EDSS protocols are merely an upper bound as we are not factoring in the need to match the frequency of the photons produced at the three single photon sources.

As we have taken the highest known quality of \textsc{cnot} implementation in this case, we now investigate the effect of lowering the success probability of the encoding and decoding operations. We first take the ratio of pair distribution rates of DED and EDSS protocols in Eqs.~\eqref{eq1} and \eqref{eq13}, which simplifies to
\begin{equation}
    \frac{R}{E[S]} = \frac{P^2 p_\mathrm{meas} L_\mathrm{wss}^2 L_\mathrm{demux}}{\eta_\mathrm{coup}^3 p_\mathrm{bsm} L_\mathrm{sw}^2},
\end{equation}
where $R$ is the rate from Eq.~\eqref{eq13} without restriction to one particular wavelength, so that $n_{\lambda_i}$ is replaced with $n$. Note that this ratio is not dependent on the distance between the nodes or the number of intermediate nodes and thus holds for any node pair (or two-node network). Using the values in Table \ref{tab:parameters}, it becomes
\begin{equation} \label{eq:ratio}
    \frac{R}{E[S]} = P^2 p_\mathrm{meas} \times 20.5482.
\end{equation}
Therefore, the pair distribution rate of EDSS1(2) matches the rate of DED when $P=0.382$ ($P=0.279$).

Averaging over all node pairs, Fig.~\ref{fig:pairDistributionRates} {\bf (b)} compares the performance of DED against EDSS for \textsc{cnot} success probability in the range $P\in[0.1, 0.5]$. We find that the number of photon pairs distributed strongly depends on the probability $P$, and the EDSS rates start to outperform DED at the values calculated using Eq.~\eqref{eq:ratio}. Therefore, if we were to use an experimental method with a typical efficiency of up to $0.25$, then DED is the optimal protocol for entanglement distribution. For $P\geq 0.279$, the EDSS-related advantage grows with $P$, which allows us to identify the regime of relatively large \textsc{cnot} success probability -- which is the most interesting one, at the experimental level, in light of the advances made very recently~\cite{Stolz2022} -- as particularly conducive of the use of such an approach.

\section{Propagation through noisy fibers} \label{sec:noise}

Imperfections in the optical fiber, such as birefringence or bending, will cause the polarization state of the photons to randomly rotate. We model this effect using Kraus operators corresponding to depolarizing noise. These operators will depend on the length $L$ of the fiber that the photon is travelling through and typical inverse length-scale $\Lambda$ within which the noise mechanism acts.

Three of the four Kraus operators are proportional to Pauli operators $K_i (\Lambda, L) = {J_i} \sqrt{1-e^{-\Lambda L}}/2$ where $\{J_1,J_2,J_3\}=\{X, Y, Z\}$ and $K_4 (\Lambda,L) = \one \sqrt{1+3 e^{- \Lambda L}}/2$. A photon in state $\rho$ undergoing depolarizing noise would become
\begin{equation}
    \rho' (\Lambda,L) = \sum_{i=1}^4 K_i(\Lambda,L) \rho K_i^\dagger (\Lambda,L).
\end{equation}
In our model, we assume that only the fibers are noisy; the photons are not exposed to any additional noise when traveling through an intermediate node. Consequently, when considering the effect of noise, each node pair is equivalent to a two-node network.

\begin{figure}[t]
    \centering
    \includegraphics[width=\columnwidth]{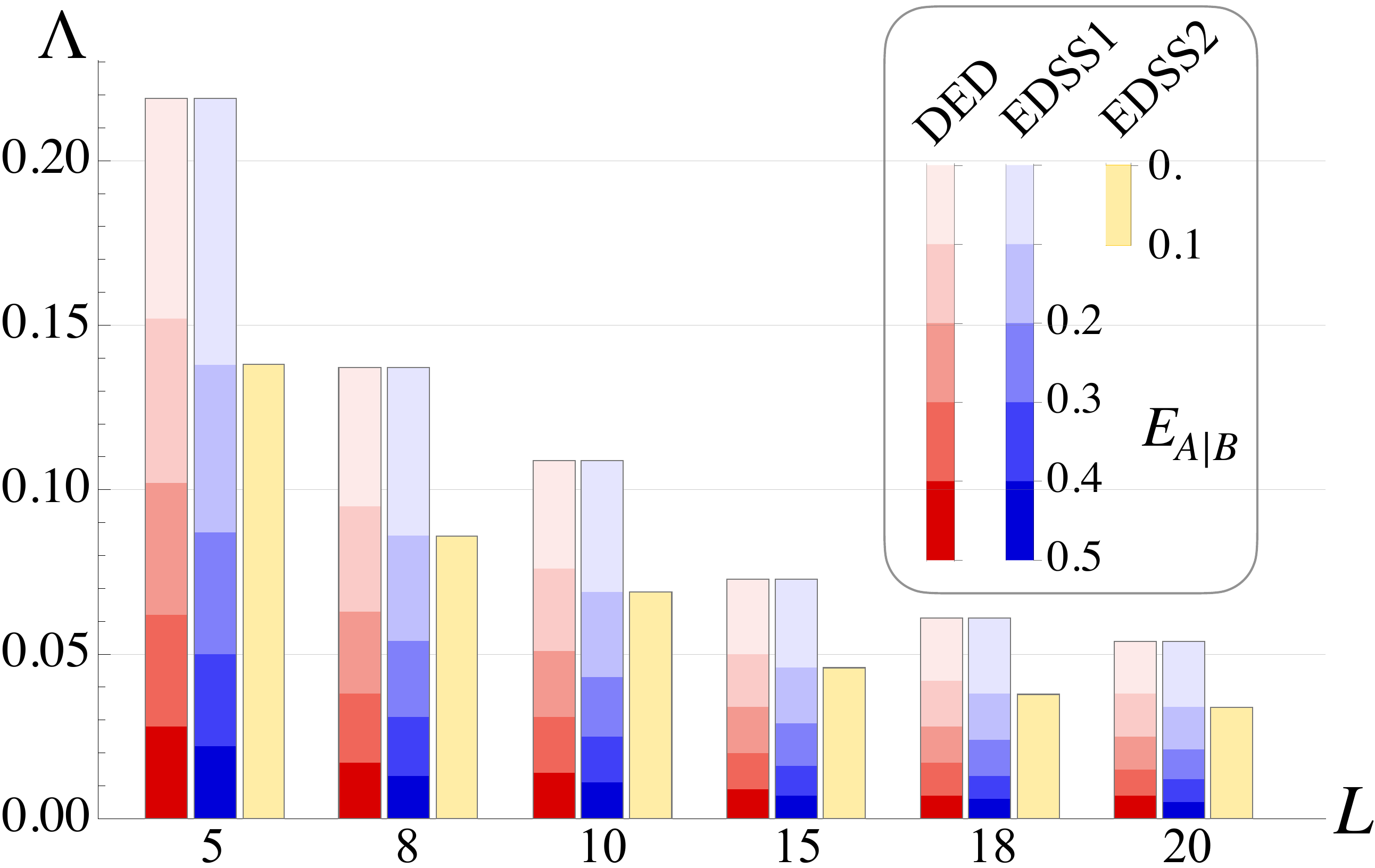}
    \caption{Degree of distributed entanglement against the distance $L$ (in units of km) between the nodes and the  depolarizing characteristic inverse length $\Lambda$. In this plot, for each sampled distance $L$, the left-most (red-colored) column is for DED, the central (blue-colored) one is for EDSS1 and the right-most (yellow-colored) one is for EDSS2.}
    \label{fig:noiseEntanglement}
\end{figure}

The amount of entanglement we can distribute as noise strength grows is shown in Fig.~\ref{fig:noiseEntanglement}. For every pair of nodes, entanglement produced using DED and EDSS1 persists under a higher level of noise than EDSS2. Moreover,  entanglement decreases more rapidly in EDSS1 than for DED, as the noise grows. We can  gather information on the respective rate by seeking the functional form that provides the best fit to the numerical values for entanglement. A simple non-linear fit approach leads to the following exponential decays 
\begin{equation}
\begin{aligned}
    f_\mathrm{DED} & \simeq 0.75 e^{- \Lambda L} - 0.25,\\
    f_\mathrm{EDSS1} &\simeq 0.64e^{-1.27 \Lambda L} - 0.16
    \end{aligned}
\end{equation}
for the DED and EDSS1 protocol, respectively, and up until $\Lambda L\simeq1.1$, when both schemes experience sudden death of entanglement~\cite{Yu2009}. Thus, quite surprisingly, DED offers a slower decay rate: as entanglement is more sensitive to noise than discord~\cite{Werlang2009}, one might think that the entanglement produced by EDSS would be the more robust of the two schemes. The reason behind such robustness is that the amount of discord communicated through EDSS1 is less than the entanglement distributed in DED. The Bell states that we send in the DED protocol have  maximum entanglement ($N=0.5$), and thus maximum discord ($D_{a|b}=1$), where we calculate discord as~\cite{Ollivier2001}
\begin{equation}
    D_{a|b} = I(a:b) - J(a|b)
\end{equation}
where $I(a:b) = S(\rho_a) + S(\rho_b) - S(\rho_{ab})$ is the quantum mutual information and
\begin{equation}
    J(a|b) = \max_{B_i^\dagger B_i} \left( S(\rho_a) - \sum_i p_i S(\rho^i_a) \right)
\end{equation}
quantifies classical correlations using a generalisation of the conditional entropy where $B_i^\dagger B_i$ is a measurement on system $b$. Entangled states distributed by DED therefore have more correlations that they can afford to lose. Meanwhile, the carrier that we send in EDSS1 only has discord $D_{ab|c}=1/3$ \footnote{Following the example in~\cite{Galve2011}, we simplify this calculation by maximizing over projective measurements on qubit $c$ instead of all positive-operator valued measures. While this is sufficient for two-qubit systems, this may not be fully accurate for three-qubit systems. Therefore, this value is in fact an upper bound of discord}. 
While this discord decays more slowly than entanglement would, EDSS1, as we have just highlighted, starts from a disadvantageous condition where quantum correlations are initially significantly less than that of DED. 

It is also worth remarking that the entangled state produced using EDSS2 is surprisingly resilient to noise. Considering the state has little entanglement to begin with, it takes relatively strong noise in the fiber to make it vanish completely; the negativity of the EDSS2 state decreases from 0.1 to 0 under the same noise strength that causes the EDSS1 state negativity to drop from 0.5 to 0.1. 

\begin{figure}[t]
    \centering
    \includegraphics[width=\columnwidth]{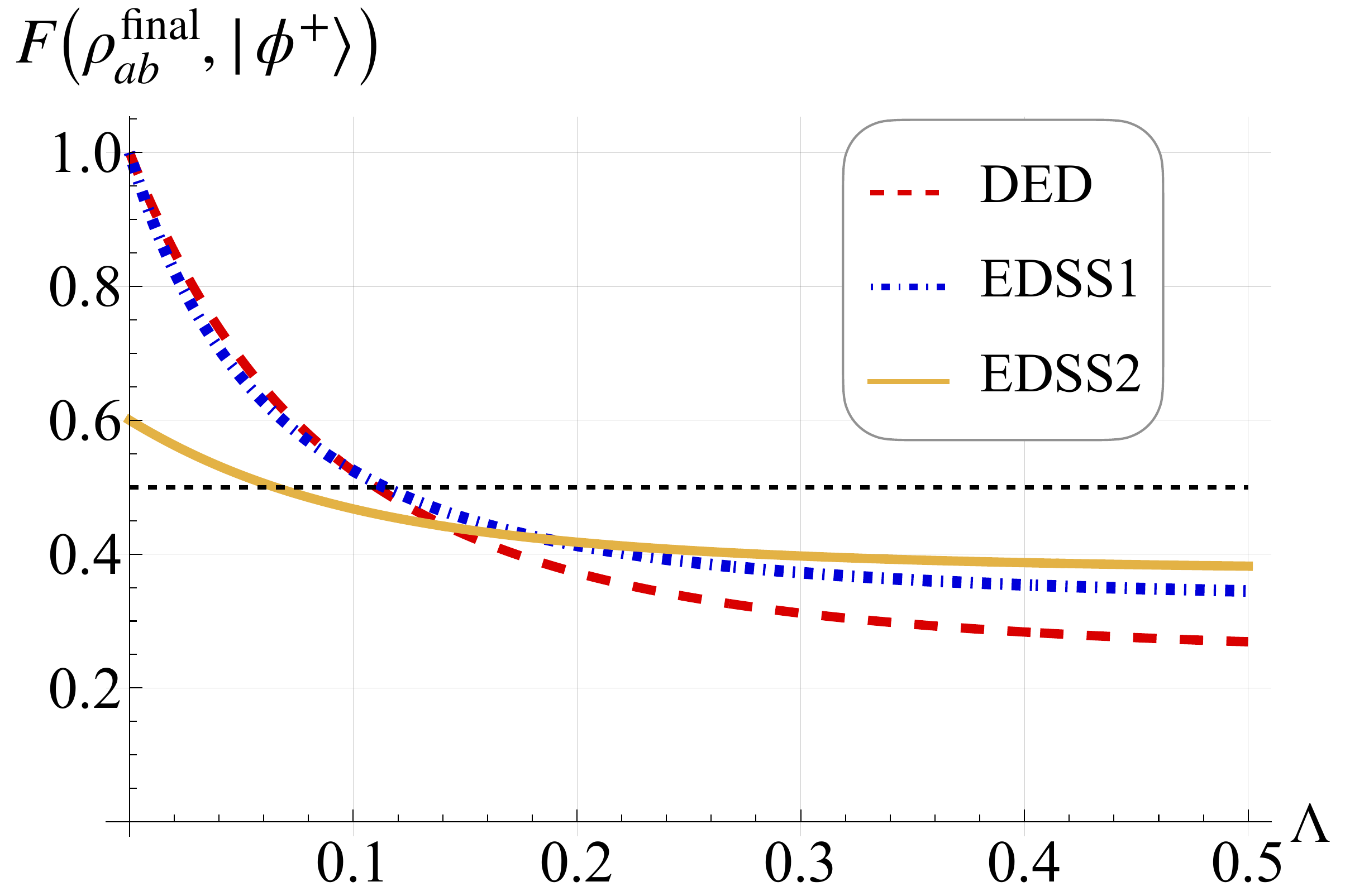}
    \caption{Fidelity between Bell state $\ket{\phi^+}$ and the state distributed, averaging over all node pairs in the network. When fidelity is less than 0.5 (black dashed line), there is no longer any entanglement in the distributed state and we are not able to distil state $\rho_{ab}^\mathrm{final}$ to form a Bell state.}
    \label{fig:noiseFidelity}
\end{figure}

Aside from entanglement, we can also consider the fidelity $F$ between the state produced via entanglement distribution $\rho_{ab}^\mathrm{final}$ and the Bell state $\ket{\phi^+}$ which we aim to generate which is calculated as
\begin{equation}
    F\left(\rho_{ab}^\mathrm{final},\ket{\phi^+}\right) = \bra{\phi^+} \rho_{ab}^\mathrm{final} \ket{\phi^+}.
\end{equation} 
In Fig.~\ref{fig:noiseFidelity}, we plot the effect of noise strength on this fidelity for each protocol. The results are averaged over all pairs of nodes as we see similar patterns for each pair. The black dashed line shows the point at which fidelity drops below 0.5; for each of the states considered (and for each node pair), the noise strength $\Lambda$ at which entanglement vanishes is the same strength needed for fidelity to fall below this threshold. 

However, as noise grows stronger, state fidelity levels out at different values: while the fidelity with a Bell state reaches a minimum of 0.25 for DED, it is at least 1/3  and 0.375 for EDSS1 and EDSS2, respectively. When exposed to very strong fiber noise therefore, we have the counter-intuitive result that the EDSS2 state is closer to a Bell state than that produced using DED.

\section{Entanglement generation rate} \label{sec:entGenerationRate}

The noisy entangled states we produce with our optical network model can be converted to maximally entangled Bell states, or ebits, using entanglement distillation protocols. In order to calculate the distribution rate of entangled pairs with fidelity above a given threshold, or how many ebits we can produce per second, we need to first find the distillable entanglement of the noisy states in Sec.~\ref{sec:noise} and then combine this with the probability of photon loss in Sec.~\ref{sec:loss}.

Note that in practice we may not need to distil the entanglement; this is dependent on how we desire to use it. There are protocols which do not need maximally entangled states~\cite{Popescu1994,Woodhead2020}; in fact, there are applications for which we need the nodes to share only discord and not entanglement~\cite{Datta2008,Dakic2012,Pirandola2014}.

To calculate the distillable entanglement, we compare several distillation protocols to discover which is optimal. The results of this process can be found in Appendix~\ref{appB}. We find that the DEJMPS protocol~\cite{Deutsch1996} gives the largest values of distillable entanglement for each state produced, no matter which entanglement distribution protocol is used or how strong the noise is in the fiber. 

To use the DEJMPS protocol, we need to start with density matrices in Bell-diagonal form
\begin{eqnarray}
    \rho = \sum_{i=1}^4 p_i \ketbra{B_i}{B_i},
\end{eqnarray}
where $\{\ket{B_i} \}_{i=1}^4=\{ \ket{\phi^+}, \ket{\phi^-}, \ket{\psi^+}, \ket{\psi^-} \}$ is the well-known Bell basis. The states produced in the three entanglement distribution protocols take this form already; otherwise, we would need to perform local operations which would convert them into this form~\cite{Preti2024}.

DEJMPS has three steps. Say we aim to distribute entanglement between nodes A and B of our network. Firstly, we rotate the photon at node A using operator $R_\mathrm{A}$ and the photon at node B with operator $R_\mathrm{B}$ where
\begin{equation} \label{eq:DEJMPSrotations}
    R_\mathrm{A} = \frac{1}{\sqrt{2}} \left( \begin{array}{cc}
        1 & -i \\
        -i & 1
    \end{array} \right), \quad R_\mathrm{B} = \frac{1}{\sqrt{2}} \left( \begin{array}{cc}
        1 & i \\
        i & 1
    \end{array} \right).
\end{equation}
These rotations can be carried out using half-wave plates and quarter-wave plates. Secondly, with two copies of the rotated entangled state distributed between nodes A and B, we perform a \textsc{cnot} operation on the two photons at node A and the two photons at node B. We must take care to choose photons from the same entangled pair to be the target. Finally, whichever pair of entangled photons was chosen to be the target is measured in the $\{\ket{H},\ket{V}\}$ basis. If we get the same outcome at each node, whether horizontal or vertical, then we have successfully increased the entanglement shared between the remaining photon pair. 

We then repeat this process until we have a two-photon state with a sufficiently high fidelity with the desired Bell state $\ket{\phi^+}$. Following the example in Ref.~\cite{Chen2023}, we take this fidelity threshold to be the demanding value of 0.998. With each iteration of the protocol, the probability that we obtain the required measurement outcomes in the final step increases.

\begin{figure}[t!]
    \centering
    \includegraphics[width=\columnwidth]{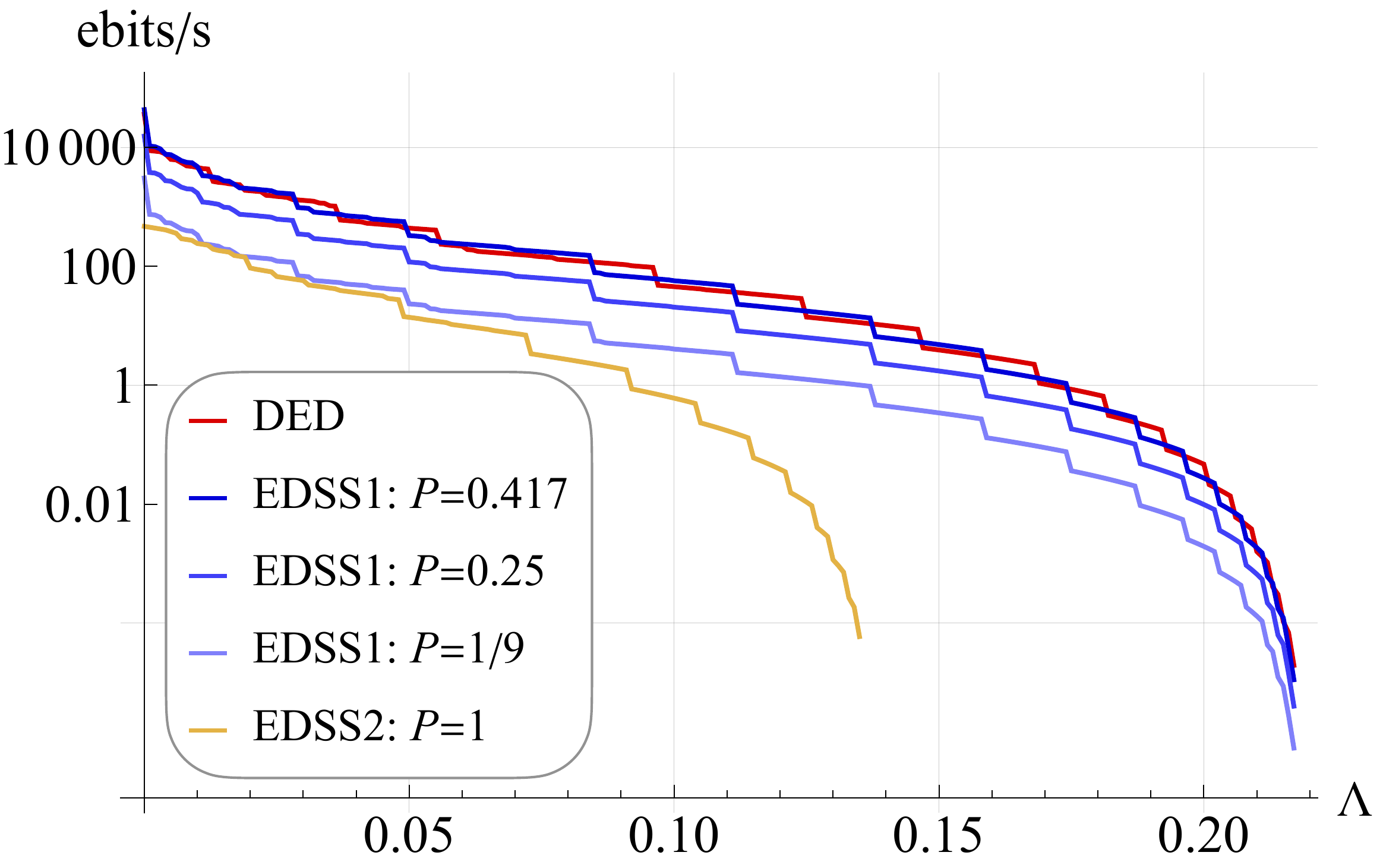}
    \caption{Pair-averaged entanglement generation rates against the rate of fiber noise. The average is calculated over all node pairs of a network.}
    \label{fig:ebitsSec}
\end{figure}

Having found the distillable entanglement, we multiply this by the probability of the protocol failing due to photon loss discussed in Sec.~\ref{sec:loss} or getting the wrong measurement outcome at the end of EDSS1 or EDSS2. It is important to note that noise has an effect on this latter probability; the stronger the depolarizing noise, the higher the probability $p_{\mathrm{meas}}^1$ of getting the desired outcome in EDSS1 and the lower the probability $p_{\mathrm{meas}}^2$ of getting the needed outcome in EDSS2. In fact, these probabilities depend on $L$ and $\Lambda$ as
\begin{equation}
    p_{\mathrm{meas}}^1 = \frac{1}{2} - \frac{e^{-\Lambda L}}{6} \quad \mbox{and} \quad p_{\mathrm{meas}}^2 = \frac{1}{2} + \frac{e^{-\Lambda L}}{8},
\end{equation}
so that they both tend towards 0.5 as the fiber gets longer and noisier.

The entanglement generation rate in our optical fiber network can be seen in Fig.~\ref{fig:ebitsSec}. As before, we average over the node pairs as the results are similar for each (see Appendix~\ref{app2nodes} for the corresponding figure for a two-node network). We see that the performance of EDSS1 depends on the success probability of the \textsc{cnot} gate; EDSS1 gives lower rates than DED for each value of $\Lambda$ when $P=0.25$ or below, but when $P=0.417$, EDSS1 is very close to DED so that the best protocol to use changes many times as $\Lambda$ grows. In fact, EDSS1 delivers the highest generation rate for $60\%$ of $\Lambda$ values in this case. The needed \textsc{cnot} success probability for EDSS1 to be optimal for over $50\%$ of values is $P=0.388$. The rate produced by EDSS2 never comes close to that of DED, even if we can perform \textsc{cphase} operations with unit probability. This is due to the lower amount of entanglement produced by EDSS2 compared to the other distribution protocols; more rounds of entanglement distillation are necessary to achieve a Bell state.

\section{Conclusions} \label{sec:conclusions}

We compared three protocols: a DED scheme based on ZALM entangled photon sources and two EDSS protocols. EDSS1 allows for the production of Bell states but requires an initial state where the carrier photon is classically correlated with the discordant photons located at remote nodes. EDSS2 is less demanding; it can be successfully carried out with a carrier qubit which is initially uncorrelated with the other systems. However, EDSS2 can only generate one fifth of the entanglement that EDSS1 and DED can.

We found that a key parameter is the probability $P$ with which we can successfully carry out an optical \textsc{cnot} or \textsc{cphase} operation in EDSS protocols. Firstly, we considered the case where we do not need to distribute maximal entanglement. In terms of vulnerability to photon loss or component failure, EDSS2 has the best performance as long as $P > 0.279$. Otherwise, DED is optimal.  

Secondly, we turned our attention to the impact of noisy fibers on the amount of entanglement distributed. Due to the reduced amount of discord present in EDSS protocols, DED is in fact the most robust protocol when faced with depolarizing noise. Entanglement produced by EDSS1 vanishes for a similar noise strength as DED, but decays at a faster rate. 

Finally, we found the rate at which Bell pairs could be generated using the three protocols. Incorporating the sensitivity to noise, the risk of photon loss, and the need for entanglement distillation procedures, we found that EDSS1 has the highest entanglement generation rate when $P>0.388$, a value which is already feasible experimentally~\cite{Stolz2022}. When we only have access to methods with lower success probabilities, DED is the best protocol. 

While in this work we consider entanglement distribution between two network nodes at a time, our design has the potential to produce entanglement between every pair of nodes simultaneously. We plan to demonstrate the routing and spectrum allocation needed to achieve this in a future work.

\acknowledgments
We acknowledge support by
the European Union’s Horizon Europe EIC-Pathfinder
project QuCoM (101046973), the Royal Society Wolfson Fellowship (RSWF/R3/183013), the UK EPSRC
(EP/T028424/1), the Department for the Economy
Northern Ireland and Science Foundation Ireland 20/US/3708, under the US-Ireland R\&D Partnership Programme.

\begin{appendix}
\section{Distributing discord}
\label{appA}

\begin{table*}[t!]
\centering
\begin{tabular}{c|c|c|c}
     & Rotation on $a$ & Rotation on $b$ & Rotation on $c$ \\ \hline
    1 & \one & \one & \one \\
    2 & $(\pi/2)$ around Z axis & $(-\pi/2)$ around Z axis & \one \\
    3 & $\pi$ around Z (Pauli matrix $\sigma_Z$) & $\pi$ around Z (Pauli matrix $\sigma_Z$) & \one \\
    4 & $(-\pi/2)$ around Z axis & $(\pi/2)$ around Z axis & \one \\
    5 & $(-\pi/2)$ around Y axis & $(-\pi/2)$ around Y axis & $\pi$ around X (Pauli matrix $\sigma_X$) \\
    6 & $(\pi/2)$ around Y axis & $(\pi/2)$ around Y axis & $\pi$ around X (Pauli matrix $\sigma_X$) 
\end{tabular}
\caption{Trilocal rotations needed to transform the initial state in Eq.~\eqref{eq:initial} to the discordant state required for EDSS in Eq.~\eqref{eq:discordState}.}
\label{tab:rotations}
\end{table*}

This Appendix addresses how to generate the necessary initial state for the EDSS protocols. For EDSS1, we can prepare the state in Eq.~\eqref{eq:discordState} using random trilocal rotations, i.e. local rotations performed on each photon which depend on a random outcome accessible to both nodes. Once the photons have been sent to the desired nodes, we start with the state 
\begin{equation} \label{eq:initial}
    \rho_\mathrm{initial} = \ketbra{D}{D}_a \otimes \ketbra{D}{D}_b \otimes \ketbra{H}{H}_c.
\end{equation} 
We are required to carry out an operation such that one of six possible trilocal rotations is performed with equal probability. Each rotation is made up of three local operations acting on each single qubit, but they are correlated so that the desired rotations are performed together. The six joint rotations are displayed in Table~\ref{tab:rotations}. 
%\AGH{The label here is wrong}. 

The initial discordant state required for EDSS2 only involves photons $a$ and $b$; photon $c$ is entirely uncorrelated with $ab$ at the beginning of the protocol. The state in Eq.~\eqref{eq:initialEDSS1} can be created in a similar way as above, but only random bilocal rotations are needed in this case. 

\section{Distilling entanglement} \label{appB}

In order to determine which distillation protocol to use for our entangled photons, we compared several recurrence protocols. We started with the two most well-known protocols: Bennett-Brassard-Popescu-Schumacher-Smolin-Wootters (BBPSSW)~\cite{Bennett1996}, the first entanglement purification protocol to be proposed, and the Deutsch-Ekert-Jozsa-Macchiavello-Popescu-Sanpera (DEJMPS) protocol~\cite{Deutsch1996}, which has been proven to be optimal for rank-3 Bell-diagonal states~\cite{Rozpedek2018} (i.e. states whose density matrix is diagonal over the Bell basis). We describe the steps of DEJMPS in the main body of the text as this protocol turned out to distil the highest amount of entanglement produced in the optical network. 

We also considered two newer and less common protocols; firstly a matter-field interaction-based protocol (which we dub as MFI)~\cite{Bernad2016a,Bernad2016b} and a \textsc{cnot}-based protocol~\cite{Preti2024} which is inspired by BBPSSW and DEJMPS, but features key differences from them.

\begin{figure*}[t!]
    \centering
    \includegraphics[width=0.8\textwidth]{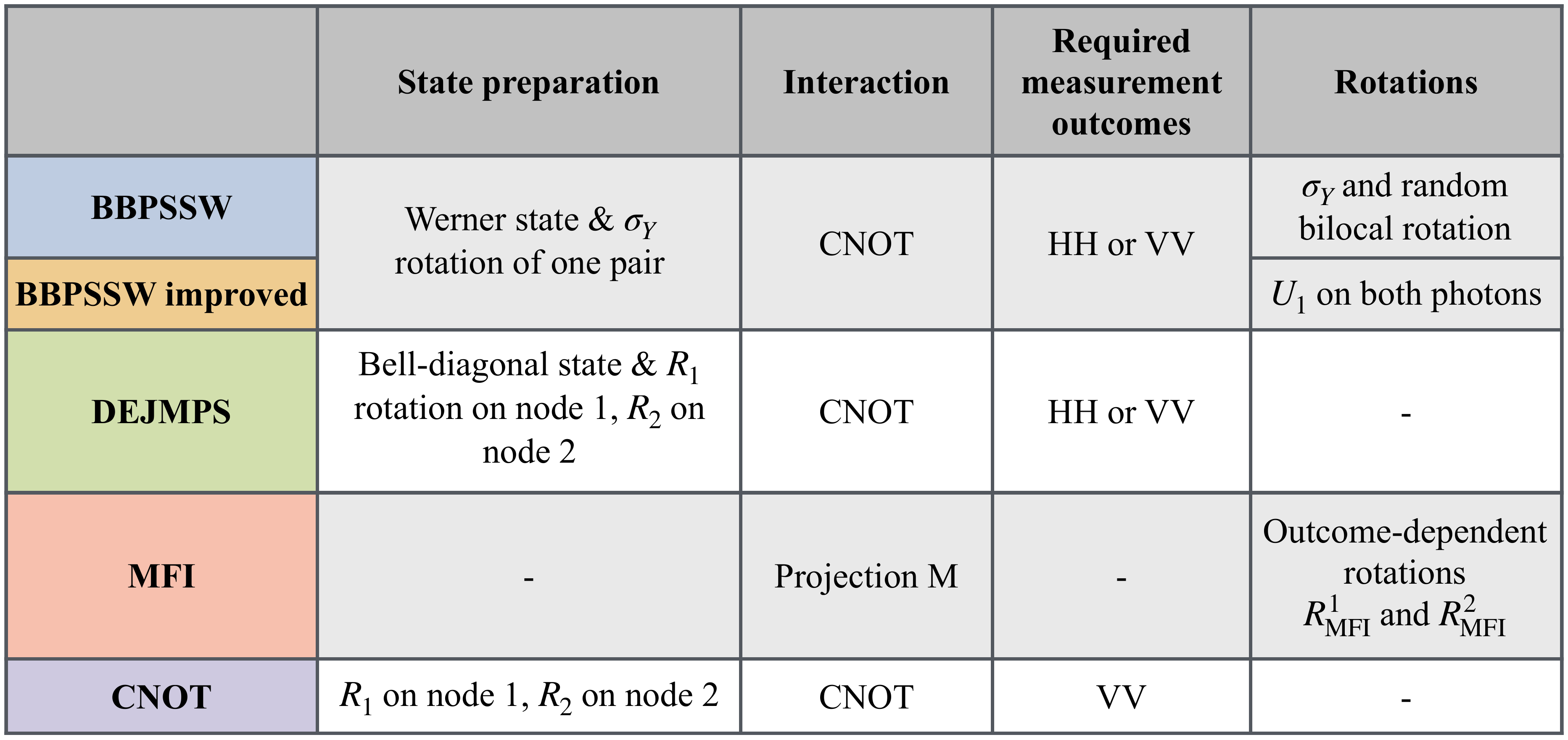}
    \caption{Summary of the iterative entanglement distillation protocols in Refs.~\cite{Bennett1996,Deutsch1996,Bernad2016a,Bernad2016b,Preti2024}.}
    \label{fig:tableDistillation}
\end{figure*}

Each protocol follows a similar pattern consisting of four parts: we begin by taking two pairs of entangled photons distributed between arbitrary nodes of our network which we label nodes 1 and 2. 
The protocols then start by implementing {\it state preparation} followed by {\it interaction} between the two photons at each node. Thirdly we perform a {\it measurement} on one pair of entangled photons in the $\{\ket{H},\ket{V}\}$ polarization basis. In all protocols except MFI, interaction is achieved using a \textsc{cnot} operation and it is the target photon pair that is measured. The BBPSSW and MFI protocols require a final {\it rotation} step. We then iterate these steps to approach a Bell state, using fidelity as a figure of merit, and setting up a demanding threshold of 0.998 as an acceptable level of similarity between the target Bell state and the evolved state of the network's nodes. A summary of each distillation protocol considered is reported in the Table in Fig.~\ref{fig:tableDistillation}. 

The first entanglement distillation protocol was BBPSSW~\cite{Bennett1996}. This protocol requires each entangled state to be transformed into a Werner state 
\begin{equation}
    W(p) = p \ketbra{\psi^-}{\psi^-} + \frac{p}{4} \one, 
\end{equation}
where $\ket{\psi^-}=\frac{1}{\sqrt{2}} ( \ket{HV}-\ket{VH})$ is a Bell state. This state can be achieved using random bilocal rotations (also called twirling operations)~\cite{Bennett1996b}. One pair of photons must then be rotated with the Pauli matrix $Y$. After interaction and measurement, the protocol is successful if the measurement outcomes at nodes 1 and 2 match. 

Finally, local rotations must be performed on the resultant photon pair. In the original protocol, this is taken to be a $Y$ rotation on just one photon, followed by a random bilocal rotation 
\begin{equation}
    \rho' = \frac{1}{3} \sum_{i=1}^3 (U_i \otimes U_i) \rho (U_i^\dagger \otimes U_i^\dagger),
\end{equation}
where $\{U_1, U_2, U_3\}$ are $\pi/2$ rotations around the $\{X,Y,Z\}$ axes in the Bloch sphere respectively. However, there is a note in Ref.~\cite{Bennett1996} about an adjustment to the final step of BBPSSW which increases the amount of distilled entanglement, namely acting on the post-measurement state with $U_1 \otimes U_1$ only. In what follows, we consider both methods, labelling the original protocol as BBPSSW and the protocol with final step changed as {\it BBPSSW improved}.  

The MFI protocol does not need the entangled states to be prepared in any particular way. Its first step is the interaction, which in this case takes the form of a measurement; we first need to project the state of the two photons at each node onto the sum of two Bell states using the joint projector
\begin{equation}
    M = \ketbra{\psi^-}{\psi^-} + \ketbra{\phi^-}{\phi^-},
\end{equation}
where $\ket{\phi^-} = \frac{1}{\sqrt{2}} (\ket{HH}-\ket{VV})$, which is only successful with limited probability. Following this, we perform the standard $\{\ket{H},\ket{V}\}$ polarization measurement on one pair of photons. Another aspect which differentiates this protocol from the previous ones is that we do not need to postselect any particular outcome; we instead perform local rotations on the remaining photons based on the outcomes. If we obtain outcome $\ket{H}$ after measuring the photon at node 1, we rotate it according to 
\begin{equation}
    R^1_\mathrm{MFI} = \left( \begin{array}{cc}
        i & 0 \\
        0 & 1
    \end{array} \right),
\end{equation}
and on obtaining outcome $\ket{V}$, we perform
\begin{equation}
    R^2_\mathrm{MFI} = \left( \begin{array}{cc}
        0 & i \\
        1 & 0
    \end{array} \right).
\end{equation}
For the photon at node 2, we do the opposite; outcome $\ket{H}$ requires rotation $R^2_\mathrm{MFI}$ and outcome $\ket{V}$ requires $R^1_\mathrm{MFI}$.

Lastly, we use the
\textsc{cnot}-based protocol~\cite{Preti2024}. This follows a similar pattern to BBPSSW and DEJMPS but has the advantage that it works on any kind of entangled state; we do not need to first prepare Werner or Bell-diagonal states. We then perform the same initial rotations as in DEJMPS in Eq.~\eqref{eq:DEJMPSrotations} with $R_1$ acting on the photons at node 1 and $R_2$ on the photons at node 2. Following interaction and measurement, the disadvantage comes with more restrictions on measurement outcomes than with BBPSSW or DEJMPS; both photons must be vertically polarised.

\begin{figure*}[t!]
    \centering
    \includegraphics[width=\textwidth]{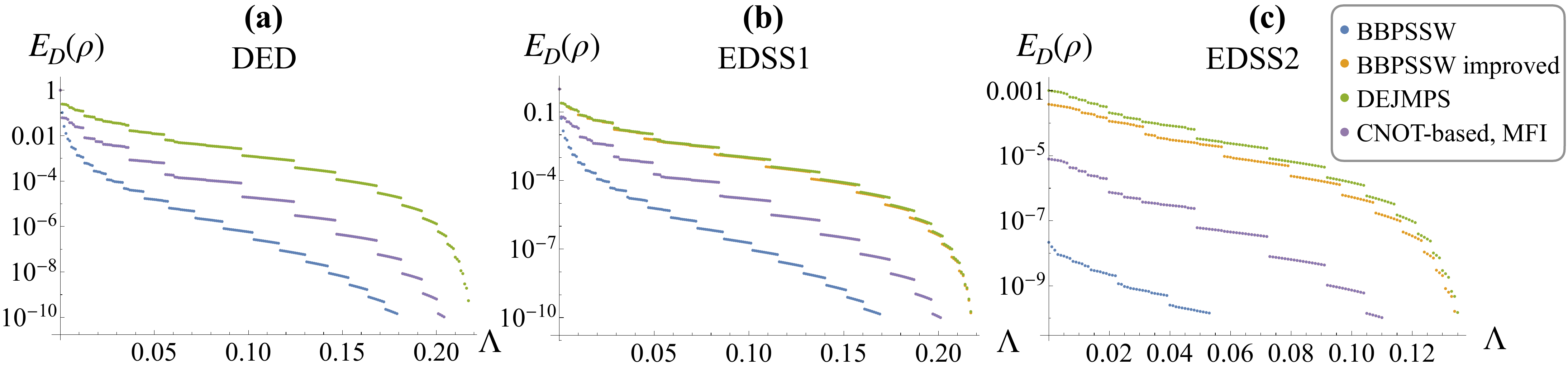}
    \caption{Distillable entanglement using five different distillation protocols as optical fiber noise grows. We take the average result for all node pairs in the network. The legend in {\bf (c)} applies to all plots. In every case, DEJMPS is the best distillation method. We do not show the results for MFI as it overlaps with the \textsc{cnot}-based method in each plot.  {\bf (a)} State produced using the DED protocol. The adjusted BBPSSW protocol and DEJMPS produce the same results. {\bf (b)} EDSS1 protocol. {\bf (c)} EDSS2 protocol. Distillable entanglement is noticeably lower than DED and EDSS1 as the state produced is not maximally entangled, even in the ideal case.}
    \label{fig:distillation}
\end{figure*}

The amount of entanglement which can be distilled using each protocol is plotted in Fig.~\ref{fig:distillation} for each entanglement distillation protocol. DED produces Bell states at the same rate when using the DEJMPS and the improved BBPSSW protocol. For both EDSS protocols, DEJMPS gives the highest generation rate, though the improved BBPSSW is very close to matching it for EDSS1.

Surprisingly, although the protocols take quite different approaches, the MFI and \textsc{cnot}-based protocols give the same output for every state regardless of how the entanglement was distributed. In every case, the original BBPSSW protocol was the least efficient.

We also considered the EPL-D protocol~\cite{Campbell2008} which, in contrast to the others, cannot be repeated. We must first perform \textsc{cnot} gates on the two photons at each node. Afterwards, we measure the photon polarization of one photon pair in the $\{\ket{H},\ket{V}\}$ basis %\AGH{$\{\ket{H},\ket{V}\}$} 
and postselect the vertical outcome. This protocol is extremely beneficial for Bell states undergoing amplitude damping noise; no matter how strong the noise, EPL-D is enough to restore the state to maximal entanglement~\cite{Campbell2010}. In other cases, EPL-D can distil more entanglement than the first iteration of other recurrence protocols; it can therefore be the optimal first step in distilling some entangled states. However, in this case of depolarizing noise, EPL-D did not bring any advantage for either DED or EDSS protocols (regardless of which Bell state was shared); in every case the entanglement of the state did not change. 

\section{Entanglement generation in a two-node network} \label{app2nodes}

In this Appendix, we provide further discussion and results on entanglement distribution across a two-node network. Instead of averaging over the node pairs of the full network illustrated in Fig.~\ref{fig:network}, we consider only nodes A and B situated 5km apart.

\begin{figure}[t!]
    \centering
    \includegraphics[width=\columnwidth]{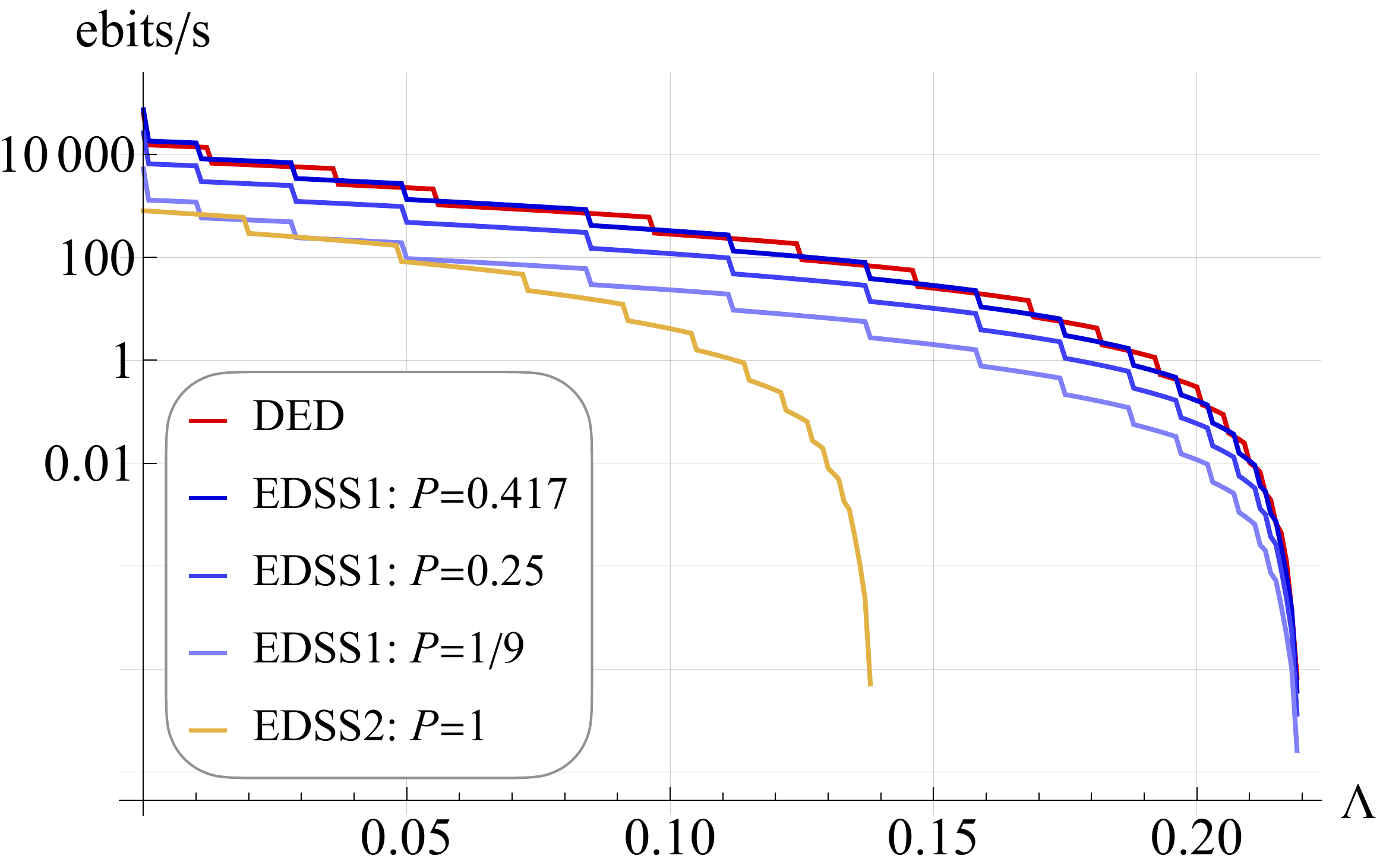}
    \caption{Entanglement generation rates against the rate of fiber noise for two nodes located 5km apart.}
    \label{fig:ebitsSecAB}
\end{figure}

The entanglement generation rates after distillation are displayed in Fig.~\ref{fig:ebitsSecAB}. The ``steps" in the plot signify the necessity of another round of distillation to achieve sufficiently high entanglement. The rate is much higher than the pair-averaged plot in Fig.~\ref{fig:ebitsSec}, due to the short distance separating A and B, as well as the lack of any lossy intermediate nodes.

Clearly there is a significant cost to performing entanglement purification. The pair distribution rates before distillation (as displayed in Fig.~\ref{fig:pairDistributionRates} {\bf (a)}) for DED, EDSS1 and EDSS2 are 61593, 73359 and 137548 pairs/s respectively when $P=0.417$. When the noise strength is just $\Lambda=0.001$, already two rounds of purification are needed for DED and EDSS1 so that their rates drop to 15245 and 18158 ebits/s. EDSS2 requires seven distillation rounds, resulting in a dramatic drop to just 137 ebits/s.

These rates are highly dependent on the amount of entanglement we need to distribute. In this instance, we have chosen a relatively high target fidelity with a Bell state of 0.998. Lowering this fidelity would raise the entanglement generation rate. It is therefore worthwhile for the network user to evaluate the amount of quantum correlations needed for the quantum protocol they want to perform; while many use Bell pairs as standard, they can nonetheless be carried out with less entangled or even non-entangled states~\cite{Dakic2012,Pirandola2014}.

\end{appendix}

\bibliography{refs.bib}

\end{document}